# The heavy quark potential in QCD with 2 flavors of dynamical quarks

U. M. Heller, Khalil M. Bitar, R. G. Edwards and A. D. Kennedy

*SCRI, Florida State University, Tallahassee, FL 32306-4052, USA*

## Abstract

We compute the heavy quark potential on configurations generated by the HEM-CGC collaboration with dynamical staggered fermions at $6/g^2 = 5.6$ and with dynamical Wilson fermions at $6/g^2 = 5.3$. The computations are done on $16^3 \times 32$ lattices, corresponding to physical sizes of about 1.6 and 2.3 fm, respectively. Up to the distances probed no sign of string breaking is detectable. We also compute the recently proposed scale $r_0$ defined by $r_0^2 F(r_0) = 1.65$.

The High Energy Monte Carlo Grand Challenge (HEMCGC) collaboration has generated ensembles of configurations with staggered sea quarks of mass $am = 0.025$ and $0.01$ at $6/g^2 = 5.6$, and with dynamical Wilson quarks for $\kappa = 0.1670$ and $0.1675$ at $6/g^2 = 5.3$, on the CM-2 at SCRI, in order to do spectroscopy and to compute some simple matrix elements [1, 2, 3]. We have used these ensembles to measure the heavy quark potential in the presence of relatively light dynamical quarks.

Comparing with the heavy quark potential obtained in the quenched approximation, we can check for effects from the dynamical quarks. Apart from an interest in its own right, such a comparison can help estimate the systematic errors from quenching in recent lattice gauge theory computations of heavy quarkonium spectra and the resulting determination of $\alpha_s$.

The heavy quark potential, or its derivative, the force $F(r)$, can also be used to obtain a scale. As recently suggested by Sommer [4] this scale does not come from the string tension, an entity that does not really exist as an asymptotic quantity in the presence of dynamical quarks, but rather from the dimensionless quantity $r_0^2 F(r_0)$ having a fixed given value. Sommer chose for this value 1.65, which leads, upon examining the phenomenological heavy quark potential, to a scale $r_0 \simeq 0.5$ fm.

The computation of the potential was fairly standard. We employed the smearing method [5] for the space-like parts of Wilson loops $W(\vec{R}, T)$ to enhance the overlap, $c_1$, with the state with lowest eigenvalue of the transfer matrix in the appropriate channel with external charges:

$$W(\vec{R}, T) = c_1 \exp\{-TV(\vec{R})\} + \cdots . \qquad (1)$$

For the construction of the larger space-like segments of the Wilson loops we also used fuzzing or blocking [6]. From the smeared Wilson loops we compute "effective" potentials

$$V_T(\vec{R}) = \log(W(\vec{R}, T)/W(\vec{R}, T+1)) \qquad (2)$$

If the overlap, $c_1$, is large the "effective" potentials $V_T(\vec{R})$ will become independent of $T$ for small $T$, before the signal is lost in the statistical noise. To illustrate the tremendous improvement smearing brings we show in Figure 1 an example of the effective potential from planar Wilson loops and from loops with smeared spatial segments as a function of $T$. As can be seen, a plateau appears with smeared loops even at the larger values of $R$, but not with the normal (unsmeared) Wilson loops.

We measured smeared Wilson loops, and hence the potential, for all vectors $\vec{R}$ lying in a plane, with the components going up to half the lattice size. The data were taken on the HEMCGC configurations with the relevant simulation parameters summarized in Table 1. More details about the configurations and the algorithms used to produce them can be found in [1, 3]. The potentials are shown in Figure 2. Rotational invariance seems to be restored quite well at distances larger than about three lattice spacings.

As can be seen, in all cases the potential appears to rise linearly at larger distances, up to the largest distances probed. It is expected that this linear rise ends (becomes screened) at some distance $r_{\text{break}}$ due to creation of a quark-antiquark pair that will bind to the external charges and thus break the string between them. Only the last two points with Wilson fermions of $\kappa = 0.1675$ hint at such a string breaking, but without statistical significance.

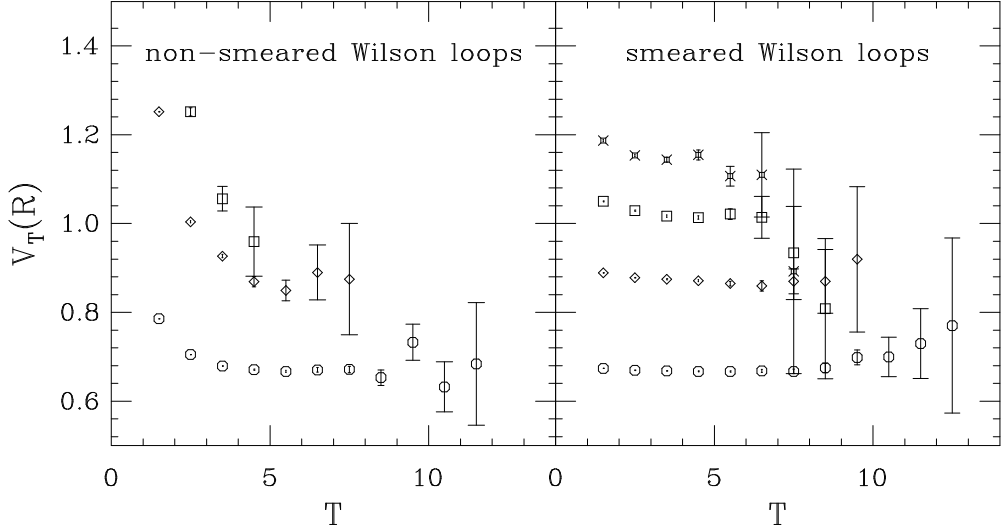

Figure 1: *The effective potential extracted from planar normal Wilson loops and from smeared Wilson loops. The sample comes from the $\beta = 5.6$, $ma = 0.025$ staggered configurations. The data are from the top for $R = 8$ ($\times$) (for which no signal was found from normal Wilson loops), 6 ($\square$), 4 ($\diamond$) and 2 ($\circ$).*

| fermions | $n_f$ | $\beta$ | $am$ or $\kappa$ | # meas | $\Delta T$ |
|---|---|---|---|---|---|
| staggered | 2 | 5.6 | 0.025 | 200 | 10 |
| staggered | 2 | 5.6 | 0.01 | 200 | 10 |
| Wilson | 2 | 5.3 | 0.1670 | 240 | 10 |
| Wilson | 2 | 5.3 | 0.1675 | 140 | 9 |

Table 1: *Summary of the configuration samples studied. $\Delta T$ denotes the number of trajectories, of unit length, between measurements.*

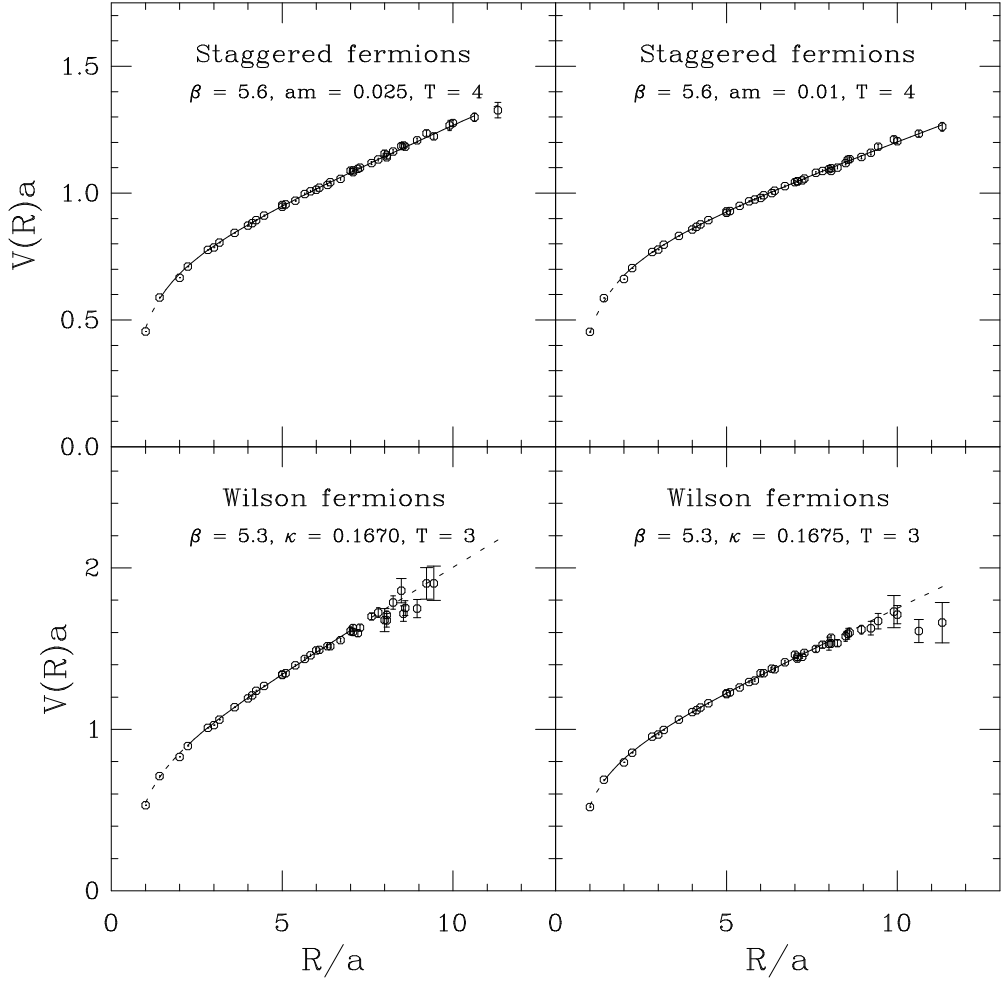

Figure 2: *The potential for the four data sets, obtained from time separations $T$ and $T+1$. The lines are the best fits, with the solid portion in the range used for the fit.*

| fermions | $am$ or $\kappa$ | T | $aV_0$ | $a^2\sigma$ | $e$ | $f$ | $r_0/a$ | $r_0\sqrt{\sigma}$ |
|---|---|---|---|---|---|---|---|---|
| staggered | 0.025 | 4 | 0.723( 2) | 0.0576( 4) | 0.312( 3) | 0.38(2) | 4.8(1) | 1.15(2) |
| | | 5 | 0.714( 5) | 0.0589( 8) | 0.302( 6) | 0.35(4) | 4.8(2) | 1.16(5) |
| | 0.01 | 4 | 0.757( 4) | 0.0481( 5) | 0.355( 7) | 0.36(3) | 5.2(2) | 1.14(4) |
| | | 5 | 0.767( 8) | 0.0470(10) | 0.373(15) | 0.29(5) | 5.2(3) | 1.13(7) |
| Wilson | 0.1670 | 3 | 0.785(14) | 0.1255(20) | 0.366(23) | 1.1(2) | 3.2(1) | 1.13(4) |
| | | 4 | 0.800(15) | 0.1209(29) | 0.378(16) | 0.61(5) | 3.2(2) | 1.11(7) |
| | 0.1675 | 3 | 0.812( 5) | 0.0976( 9) | 0.382( 6) | 0.57(4) | 3.7(2) | 1.16(6) |
| | | 4 | 0.805(11) | 0.0978(22) | 0.374(12) | 0.59(6) | 3.6(2) | 1.13(6) |

Table 2: *Summary of results from fits to the effective potentials using Eq. (3) on the four data ensembles. The last two columns give the scale $r_0/a$ determined from $r_0^2 F(r_0) = 1.65$ and the dimensionless quantity $r_0\sqrt{\sigma}$.*

A rough estimate of the string breaking distance, $r_{\text{break}}$, would be $r_{\text{break}}\sigma = 2m_q$, where $\sigma$ is the string tension and $m_q$ a light constituent quark mass, e.g. $m_q = m_\rho/2$. This would lead to the estimate $r_{\text{break}} \approx 0.8$ fm, which is somewhat shorter than the largest distances probed. Alternatively, one can use the results from the light spectroscopy [1, 3] for $m_\rho$ and the string tension determined below: this leads to the estimate $r_{\text{break}}/a \approx 11$ and 5.5 for staggered and Wilson sea quarks respectively. For Wilson sea quarks we can measure the potential for somewhat larger distances than the rough estimate for $r_{\text{break}}$ and it appears that the string breaking sets in, at the earliest, at almost twice the rough estimate.

Since we do not observe a string breaking we used the same ansatz that is common in quenched simulations [7] for fitting the potential

$$V(\vec{R}) = V_0 + \sigma R - \frac{e}{R} - f\left(G_L(\vec{R}) - \frac{1}{R}\right) \quad , \tag{3}$$

where $G_L$ denotes the lattice Coulomb potential. The last term takes account of the lattice artefacts present at small distances; it helps in getting good fits that also include rather small distances. We used fully correlated fits with the covariance matrix estimated by a bootstrap method. In all cases, the best fit values obtained in this way did not differ significantly from those of naive, uncorrelated fits. The best fits, selected according to the criterion of having maximal "quality", defined as the product of confidence level times the number of degrees of freedom divided by the relative error of the string tension, are also shown in Figure 2. The fit parameters, in each case for two choices of the distance $T$ from which the "effective" potential was taken, are listed in Table 2.

From the potential we can compute the force between static quarks by taking appropriate differences of the potential

$$F_j(\tilde{r}) = V(\tilde{r}) - V(\tilde{r} - \vec{e}_j) \tag{4}$$

Recall that we computed the potentials for $\vec{r}$ in a plane, so we get two non-vanishing components of the force. We are interested in its magnitude $F(\tilde{r}) = \sqrt{F_1^2(\tilde{r}) + F_2^2(\tilde{r})}$. For the

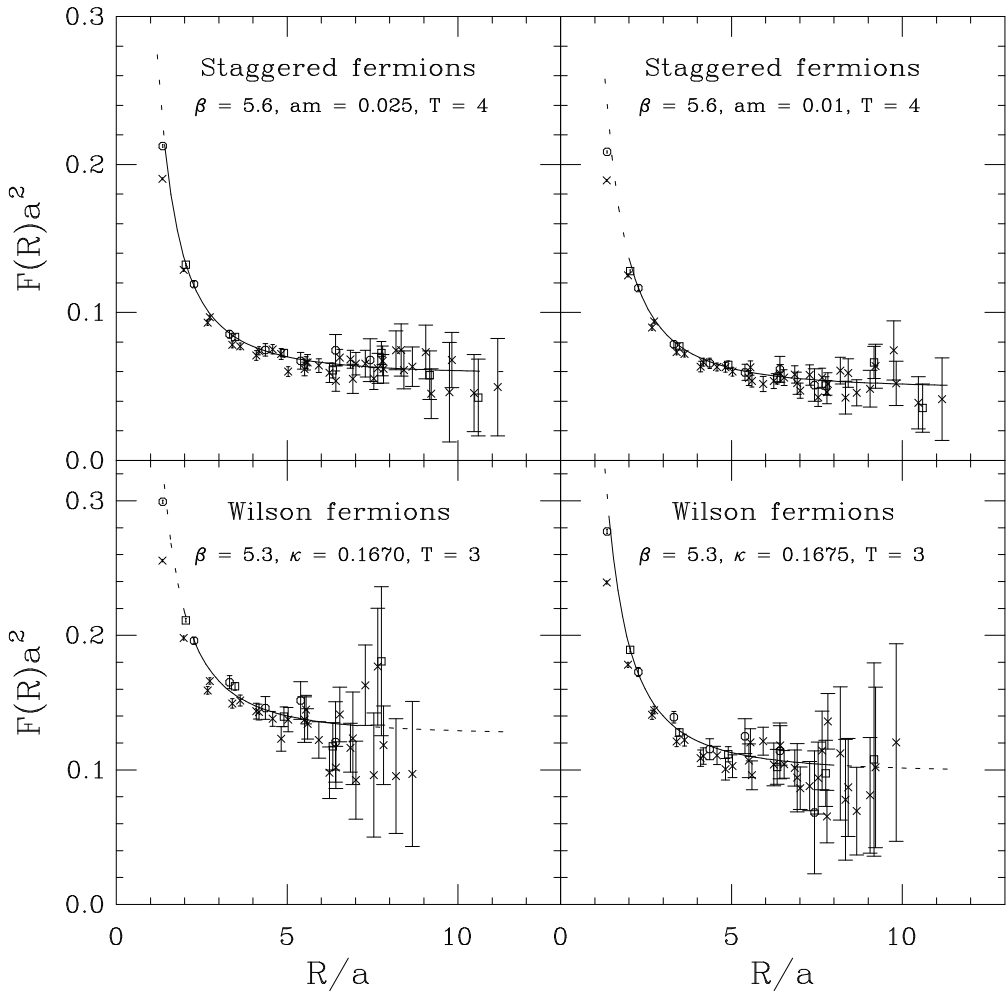

Figure 3: *The force for the four data sets, obtained from taking differences of the potential shown in Figure 2 as described in the text. The line is the derivative of the best fit to the potential with f set to zero.*

distance $\tilde{r}$ we choose

$$\tilde{r} = F_L^{-1/2}(\tilde{r}) \qquad (5)$$

where $F_L(\tilde{r})$ is computed as in Eq. (4) but using the lattice Coulomb potential $G_L$ instead of $V(r)$. With this choice $F(\tilde{r})$ is a tree-level improved observable. The force obtained in this way is shown as crosses in Figure 3.

If we know the potential on a ray $n\vec{d}$, we can compute the force along this ray directly by

$$F(\tilde{r}) = \left[V(n\vec{d}) - V((n-1)\vec{d})\right] |\vec{d}|^{-1} \qquad (6)$$

where $\tilde{r}$ is again computed from the lattice Coulomb potential. The force computed in this way from on-axis potentials is shown as octagons and from the potential along a diagonal as squares in Figure 3. The force, obtained from taking differences, is much noisier than the potential, and rotational invariance is not obeyed quite as well.

| fermions | $\beta$ | $am$ or $\kappa$ | $a(r_0)$ | $a(m_\rho)$ | $a(m_P)$ | $\sqrt{\sigma}$ |
|---|---|---|---|---|---|---|
| staggered | 5.6 | 0.025 | 0.102( 2) fm | | | 454( 8) MeV |
| | 5.6 | 0.01 | 0.096( 4) fm | | | 450(16) MeV |
| | 5.6 | 0.0 | 0.092( 7) fm | 0.110( 2) fm | 0.119( 2) fm | |
| Wilson | 5.3 | 0.1670 | 0.156( 9) fm | | | 446(16) MeV |
| | 5.3 | 0.1675 | 0.135( 7) fm | | | 458(24) MeV |
| | 5.3 | $\kappa_c$ | 0.116(16) fm | 0.109( 2) fm | 0.124( 3) fm | |

Table 3: *The lattice spacing from $r_0$ for the four ensembles, the lattice spacing from $r_0$, $\rho$ and nucleon mass in the extrapolation to zero quark mass or $\kappa_c$, and $\sqrt{\sigma}$ with the scale set by $r_0$.*

As suggested by Sommer [4] the force can be used to obtain a scale which determines the lattice spacing. He suggested to define $r_0$ from the dimensionless quantity

$$r_0^2 F(r_0) = 1.65 \qquad (7)$$

and found from phenomenological heavy quark potentials that $r_0 \simeq 0.5$ fm. There are several advantages in using $r_0$ to set the scale. It is well defined even in the presence of dynamical quarks where an asymptotic string tension does not exist. It does not require any fitting but only some interpolation of the force, and hence avoids biases from a particular fitting procedure. It comes from a relatively small distance where measurements are usually more accurate. The drawback comes from having to take numerical derivatives, which increases the noise.

The values obtained for $r_0/a$ and for the dimensionless combination $r_0\sqrt{\sigma}$ are shown in the last two columns of Table 2. The combination $r_0\sqrt{\sigma}$ does not seem to depend on the kind of sea quarks present, or their mass, within about 5% statistical errors. It also seems to agree with the results from quenched simulations, compiled in Ref. [8], again within errors.

We compare, in Table 3, the lattice spacing extracted from the scale $r_0 = 0.5$ fm, with the lattice spacing extracted from light hadron masses from the extrapolation to zero quark mass [1] or to $\kappa_c$ [3]. We also list the value for the string tension, obtained from using $r_0$ to set the scale. For Wilson sea quarks the lattice spacings from $r_0$ and from light hadron spectroscopy are in reasonable agreement but the determination from $r_0$ has large errors from the extrapolation to $\kappa_c$. For staggered sea quarks the lattice spacing from $r_0$ is considerably smaller than from the light staggered spectroscopy. However, the HEMCGC collaboration has also done spectroscopy with Wilson valence quarks on the ensembles with staggered dynamical fermions [2] and found lattice spacings, from the rho mass, of $a \simeq 0.099$ fm and 0.092 fm for quark masses $am = 0.025$ and 0.01. These values, and their extrapolation to $am = 0$, $a \simeq 0.088$ fm, agree better with the lattice spacings determined from $r_0$.

In conclusion, we have measured the heavy quark potential in the presence of two flavors of moderately light sea quarks, both staggered and Wilson. Up to the largest distances probed, about 1.5 fm for the Wilson quarks, we did not see any sign of string breaking.

Hence this string breaking seems to occur at the earliest at about twice the distance one would expect from a naive rough estimate of 0.8 fm. Qualitatively, the potentials look very similar to quenched potentials at comparable lattice spacings, as has already be seen in [9] for the case of four flavors of staggered sea quarks. Also the dimensionless quantity $r_0\sqrt{\sigma}$, where $r_0$ is determined from $r_0^2 F(r_0) = 1.65$, seems to agree well with quenched results. The Coulomb coefficient, $e$, in our fits, eq. (3), is of comparable magnitude to those obtained in quenched calculations. It would be nice to better determine the running coupling constant in the presence of dynamical fermions, since there one should see a difference compared to quenched calculations [10]. That would require simulations at smaller lattice spacing, to get good resolution of the short-distance part of the heavy quark potential.

This research was supported in part by the DOE under grants # DE-FG05-85ER250000 and # DE-FG05-92ER40742. The computations have been performed on the CM-2 at SCRI. The use of the HEMCGC configurations is gratefully acknowledged. UMH would like to thank Jochen Fingberg for the routines used to fit the potential and discussions about the fitting procedure. He also thanks Tom DeGrand and Bob Sugar for discussions and a critical reading of the manuscript.